\documentclass[a4paper,11pt]{article}
\usepackage{pos}

\title{Fermionic Neural Networks through the lens of Group Theory}

\author*[a, b]{J. Rozalén Sarmiento}
\author[a,b]{A. Rios}

\affiliation[a]{Dept. Física Quàntica i Astrofísica, Universitat de Barcelona,\\
  Carrer Martí i Franquès 1, 08028 Barcelona, Spain}

\affiliation[b]{Institut de Ciències del Cosmos, Universitat de Barcelona,\\
  Carrer Martí i Franquès 1, 08028 Barcelona, Spain}

\emailAdd{javi.rozalen@icc.ub.edu}
\emailAdd{arnau.rios@icc.ub.edu}

\abstract{
We present an overview of the method of Neural Quantum States applied to the many-body problem of atomic nuclei. Through the lens of group representation theory, we focus on the problem of constructing neural-network ansätze that respect physical symmetries. We explicitly prove 
that determinants, which are among the most common methods to build antisymmetric neural-network wave functions, can be understood as the result of a group convolution. We also identify the reason why this construction is so efficient in practice compared to other group convolutional operations. We conclude that group representation theory is a promising avenue to incorporate explicitly symmetries in Neural Quantum States.

}

\FullConference{10th International Conference on Quarks and Nuclear Physics (QNP2024)\\
8-12 July, 2024\\
Barcelona, Spain\\}

\begin{document}
\maketitle

\section{Introduction}
A key goal of nuclear structure theory is to obtain the ground-state wave function of atomic nuclei starting from first principle simulations. A variety of approximations and computational methods have been proposed throughout the years to solve the nuclear many-body problem starting from phase-shift-equivalent interactions~\cite{hergert_2020_ab_initio}. 
In 2017, a novel method, dubbed Neural Quantum States (NQS), was put forward by Carleo and Troyer~\cite{carleo2017solving}. In this approach, neural networks act as wave function ansätze and the energy is the cost function of the system. This reinforcement learning approach is akin to a variational Monte Carlo (VMC) setting in quantum mechanics. Since its inception, NQS have been proven to yield competitive results in a variety of quantum many-body settings, including quantum chemistry~\cite{pfau2020ab} and nuclear structure~\cite{Keeble2020,Adams2021,Gnech2023}.




Currently, simulations indicate that NQS can become a competitive method to perform \textit{ab initio} simulations even in the medium mass region of the nuclear chart~\cite{Gnech2023}. The success of the  method lies on two heuristic ideas. On the one hand, neural networks (NN) are very expressive ansätze. On the other, empirical evidence indicates that the information is efficiently encapsulated in the network parameters, allowing for energy minimisations that are polynomial in particle number. Here, we specifically look at the first point. We discuss methods to incorporate relevant physical symmetries in NQS, so that ansätze are restricted to the correct subspaces. In turn, this may affect the second point, in that information is encoded more efficiently in a symmetry-preserving form, thus leading to more efficient energy minimisation. 

\section{Symmetries in Neural Quantum States}
When building wave function ansätze of a physical system, we may want to impose some information about the system \textit{a priori}. In the case of nuclei, for instance, protons and neutrons are the relevant degrees of freedom. Since these are fermions, the corresponding wave function should be totally antisymmetric with respect to the exchange of particle coordinates. Other relevant symmetries that may be explicitly built into the ansatz involve the spin, isospin or rotational degrees of freedom.  

The specific way in which the symmetries of a system dictate the transformation properties of our wave functions can be systematically studied using group representation theory~\cite{wu_ki_tung_group_theory}. 
Let us consider a system governed by a Hamiltonian $\hat H$, and an operator, $\hat P$, that commutes with it, $\left[ \hat H, \hat P\right] = 0$. At this point, we know that $\hat P$ is related to a symmetry of the system. For instance, expectation values of $\hat P$ built with eigenstates of $\hat H$ are constants of motion. 

The first version of Schur's lemma applied in this context indicates that: \textit{Let $\rho$ be an irreducible representation (irrep) of an abelian group $G$ on a vector space $V$, and $A$ be an arbitrary operator on $V$. If $\rho(g)A = A\rho(g) \, \forall g \in G$, then $A$ is a multiple of the identity operator, $A=E\mathbb{I}$.}
In other words, provided an irreducible representation of $\hat P$, $\rho_V$, and constraining $\hat H$ to the irreducible invariant subspace where $\rho_V$ acts, $\hat H_V$, $\rho_V$ commutes with $\hat H_V$\footnote{Notice that $\hat H_V$ can be computed by projection of $\hat H$ onto the relevant subspace $V$.}. By Schur's lemma, we know that $\hat H_V$ must act as a multiple of the identity in this subspace, and we can identify that the scalar multiple $E$ is the energy of the states living in this subspace. In this setting, the component of the wave function that lives in $V$ must transform under $\hat P$ by the irrep $\rho_V$.

For example, let us consider a single particle in a one-dimensional harmonic oscillator well. There, it is evident that $\left[ \hat H, \hat \Pi \right]=0$, where $\hat \Pi$ is the parity operator, which realizes the action of the group of order $2$. This immediately tells us that the eigenstates of $\hat H$ are split into two sets: those transforming with (I) the trivial representation, $\phi(-x) = \hat \Pi \phi(x) = \phi(x)$, and with (II) the alternating representation, $\phi(-x) = \hat \Pi \phi(x) = -\phi(-x)$. This coincides with the common knowledge that the eigenstates of the harmonic oscillator are either odd or even under reflection. 

While Schur's lemma clarifies the nature of symmetries and groups, it does not provide a constructive approach to incorporate symmetries into NN architectures. 
For simple systems, symmetries may be incorporated employing intuition. For instance, in the harmonic oscillator example above, we can achieve both types of transformations by (anti-)symmetrizing the output of a neural network, $\phi(x)$, as follows \cite{mattheakis2020physicalsymmetriesembeddedneural}:
\begin{equation}
    \psi^\pm(x) = \phi(x) \pm \phi(-x) \, .
\end{equation}
As symmetries and their corresponding operators become more complex, or as the number of symmetries increases, one would like to resort to a systematic approach to generate NN architectures. 

Group-convolutional neural networks (G-CNNs) were introduced as a generalization of the standard convolutional neural networks under the action of any group \cite{cohen2016groupequivariantconvolutionalnetworks}. These approaches were originally motivated by image processing, where translation equivariance is built into the network through convolutional layers,
\begin{equation}
    \phi_l(f_{l-1})(x_1, x_2) = \sum_{\vec{u}=(u_1, u_2)} f_{l-1}(x_1 - u_1, x_2 - u_2) k_l(u_1, u_2) \, ,
\end{equation}
where $f_{l-1}$ is the output of the $(l-1)-$th layer; $k_l$, is the kernel in the $l-$th layer, and $\phi_l$ is the $l-$th layer of the network. The vector $\vec{u}$ is an element of the group of translations in two dimensions, $T(2)$. In analogy to this expression, we can envisage a generalization of the construction to any group $G$:
\begin{equation}{\label{eq:group-convolution-regular}}
    \phi_l(f_{l-1})(x) = \sum_{g\in G} f_{l-1}(g^{-1}x) k_l(g) \, .
\end{equation}
One can prove that this construction leads to the equivariance of $\phi_l$ under transformations in $G$~\cite{cohen2016groupequivariantconvolutionalnetworks}. One can also show that, under some minimal constraints, the converse also holds, this is: \textit{any equivariant layer must be group-convolutional} \cite{kondor2018generalizationequivarianceconvolutionneural, cohen2020bigpaper}. 

\section{Fermionic networks are also convolutional}
To see how this can be put into practice in nuclear theory applications, we consider fermionic systems. Historically, antisymetry of a wave function has been achieved by using determinants~\cite{pfau2020ab}, or some generalizations of these, such as Pfaffians~\cite{Fore2023}. This choice is usually motivated in terms of physical intuition.
We now provide a group-theoretical approach to justify this choice, indicating that determinants (or generalizations thereof) arise as a special case of group convolutions.

What type of equivariance is required in the fermionic case?
The Pauli exclusion principle tells us that, for systems of identical fermions, a permutation $\sigma$ of particle inputs lead to the same wave function up to a sign,
\begin{equation}{\label{eq:anti-symmetry}}
    \psi(x_{\sigma(1)}, x_{\sigma(2)}, \dots, x_{\sigma(N)}) = (-1)^P \psi(x_1, x_2, \dots, x_N) \, .
\end{equation}
$(-1)^P$ is the so-called parity of the permutation and $x_N$ represents a  particle coordinate (that could be position, but may also include spin or other degrees of freedom). 
The symmetry group of relevance here is the symmetric group acting on $N$ elements, $S_N$. The permutation of particle indices (left-hand side of Eq.~(\ref{eq:anti-symmetry})) is equivalent to the action of the regular representation of $S_N$ onto a vector containing all particle positions. 
The change of sign introduced to the wave function (right-hand side of Eq.~(\ref{eq:anti-symmetry})) corresponds to the action of the \textit{alternating} irreducible representation of $S_N$. We note that the derivation of the irreducible representations of $S_N$ is not trivial, but possible~\cite{wu_ki_tung_group_theory}.

A map that connects these two representations is known in group representation theory as an \textit{intertwiner}. Formally, an intertwiner is a linear map $\Psi$ such that
\begin{equation}
    \rho(g) \Psi = \Psi \pi(g) \quad \forall g\in G \, ,
\end{equation}
where $\rho, \pi$ are representations of $G$ acting on the output and input spaces of $\Psi$, respectively. In a G-CNN setting, one may use an intertwiner as the linear operator in the equivariant layers, thus preserving equivariance on a layer-by-layer basis. 
Cohen \emph{et al.}~demonstrated that 
such an intertwiner can always be expressed as a group convolution \cite{cohen2016steerablecnns, cohen2020bigpaper}. 
One can prove that the operator 
\begin{equation}
    \Psi = \frac{1}{\vert G \vert} \sum_{g\in G} \rho(g) A \pi(g^{-1})
\end{equation}
intertwines the two representations $\rho$ and $\pi$. Here, $A$ is any linear map between the vector spaces onto which the representations act, $A:V_\pi \longrightarrow V_\rho$. 

We can now apply this technique to the specific case of fermionic antisymmetry. Using $G=S_N$, and replacing $\rho$ by the alternating irrep, we find:
\begin{align}{\label{eq:intertwiner-determinant}}
    \Psi = \frac{1}{N!} \sum_{\sigma \in S_N} (-1)^P A \pi(\sigma^{-1}) \, .
\end{align}
We now show that Eq.~(\ref{eq:intertwiner-determinant}) reduces to a Slater determinant. 
Let us use the label $l$ for the last layer of the network. The usual choice is to build the $(l-1)-$th layer such that its output lies in $\mathbb{R}^N$ \cite{pfau2020ab, keeble2023machine}. 
In the simplest case, each component of the vector in $\mathbb{R}^N$ is a single-particle orbital, and we can embed the vector of $N$ orbitals into the tensor space $\mathcal{H}_1 \otimes \mathcal{H}_2 \otimes \dots \otimes \mathcal{H}_N$. 
A natural realization of $S_N$ on this tensor product space is
\begin{equation}
    \pi(\sigma) \phi^{l-1}(\vec{x})= \pi(\sigma) \left[ \phi_1(x_1)\otimes \phi_2(x_2) \otimes \dots \otimes \phi_N(x_N) \right] = \phi_1(x_{\sigma(1)}) \otimes \phi_2(x_{\sigma(2)}) \otimes \dots \otimes \phi_N(x_{\sigma(N)}) \, .
\end{equation}
Using this realization as the $\pi$ representation in Eq.~(\ref{eq:intertwiner-determinant}) and using $A=\mathbb{I}$, we find the usual definition of the Slater determinant,\footnote{The usual definition of the Slater determinant has an extra prefactor $\sqrt{N!}$, but this does not affect the equivariance property of our map.}
\begin{equation}{\label{eq:slater-determinant}}
    \Psi \phi^{l-1}(\vec{x}) = \frac{1}{N!} \sum_{\sigma \in S_N} (-1)^P \phi_1(x_{\sigma^{-1}(1)}) \otimes \phi_2(x_{\sigma^{-1}(2)}) \otimes \dots \otimes \phi_N(x_{\sigma^{-1}(N)}) \, .
\end{equation}

The demonstration above assumes that the vector components are single-particle orbitals. In general, however, the components of the $(l-1)-$th layer, $\phi^{l-1}(\vec{x})$, need not be single-particle orbitals
so long as they transform equivariantly with respect to particle exchanges~\cite{pfau2020ab,keeble2023machine}. This can be achieved by employing backflow-enhanced orbitals, $\phi(x_i;\{x_{/i}\})$, where $\{x_{/i}\}$ represents a permutation-invariant feature of all coordinates, except $x_i$. These backflow correlations usually improve the expressivity of the ansatz~\cite{pfau2020ab,Fore2023,keeble2023machine}. Pfaffians have also been proposed as a tool to build antisymmetric NNs \cite{Fore2023}. In our setting, they correspond to the case where $\phi^{l-1}$ is the product of a single pairing orbital calculated at different pairs of coordinates. Defining $\pi$ as the regular representation, and applying $\Psi$ to $\phi^{l-1}$, one obtains the usual expression for the Pfaffian.

One may naively expect that the sum over the group $S_N$ in Eq.~(\ref{eq:slater-determinant}) would involve a time complexity that is proportional to the number of terms in the sum, $\mathcal{O}(N!)$. However, Eq.~(\ref{eq:slater-determinant}) is clearly also the Leibniz formula of the determinant. 
We can thus leverage the usual algorithms with $\mathcal{O}(N^3)$ time complexity to compute these determinants.
This might be a special feature of the alternating irrep of $S_N$, in the sense that other irreps or groups might not exhibit representations which allow for such a reduction in the time complexity of convolutional operations.

In the construction outlined above, physical (anti-)symmetry has dictated the type of equivariance of the last layer of the network. 
Group-theoretical arguments might not result extremely useful to build equivariance in the last layer, provided one already has an intuition on how to express it mathematically. However, we might benefit from such arguments in the intermediate layers, where a systematic, yet flexible, approach that remains equivariant across layers is desirable. 
In fact, there is some freedom in choosing the type of equivariance of the intermediate layers. For every layer, one could repeat a construction based on intertwiners, but this time, between two representations of our choice. 
There is no unique answer on which representations to choose, but the authors of Ref.~\cite{cohen2016steerablecnns} suggested a systematic method. Every group representation can be decomposed into a direct sum of irreducible representations. From this point of view, the difference between any two representations lies in the number of times each irrep is used in the construction. In this sense, the choice of representations across layers may reduce to selecting a sequence of natural numbers at every layer. This layer-by-layer construction may lead to an expressive, yet symmetry-restricted representation, which may ultimately provide more powerful NN architectures.

\section{Conclusions and future outlook}


The application of the NQS method to nuclear theory may benefit substantially from a solid understanding on how to incorporate nuclear symmetries into wave function ansätze. A promising approach in this direction is given by G-CNNs, which provide a clear constructive strategy to generate equivariant neural network architectures. We have discussed an illustrative example in terms of fermionic antisymmetry, and shown that the G-CNN strategy leads to an architecture that is akin to Slater or backflow-enhanced determinants. 

We foresee several applications of the methods discussed here. 
In the context of fermionic symmetry, it would be interesting to investigate how the usual constructions of antisymmetry (i.e., Slater determinants, neural backflows...) arise from the earlier layers of the NNs.
In a longer term, and thinking about nuclear physics needs, we would also like to extend this formalism to incorporate equivariance not only under one group (say, antisymmetry), but under more than one (e.g. spin and isospin). Ultimately, this may pave the way to generic NQS ansätze that can capture several symmetries simultaneously, yielding more cost-effective minimisation problems. 

\section{Acknowledgments}
This work is financially supported by 
MCIN/AEI/10.13039/501100011033 from grants: 
PID2020-118758GB-I00 and PID2023-147112NB-C22; 
RYC2018-026072 through the “Ram\'on
y Cajal” program funded by FSE “El FSE invierte en tu futuro”; 
CNS2022-135529 through the “European Union NextGenerationEU/PRTR”; 
CEX2019-000918-M through the “Unit of Excellence Mar\'ia de Maeztu 2020-2023” award to the Institute of Cosmos Sciences; and by the Generalitat de Catalunya, grant 2021SGR01095. J. R. S. acknowledges the support from the predoctoral program AGAUR-FI ajuts (2024 FI-1 00824) Joan Oró de la Secretaria d'Universitats i Recerca del Departament de Recerca i Universitats de la Generalitat de Catalunya i del Fons Europeu Social Plus.

\bibliographystyle{JHEP} 
\bibliography{bibliography}


\end{document}